\documentclass{article}
\usepackage{amssymb}
\usepackage{epsfig}
\usepackage{graphicx}
\usepackage{amsmath}
\usepackage{epigraph}

\begin{document}

\title{Geometric evolution of the Reynolds stress tensor}
\date{\empty }
\author{S. GAVRILYUK\thanks{
University of Aix-Marseille \& C.N.R.S. U.M.R. 6595, IUSTI, 5 rue E. Fermi,
13453 Marseille Cedex 13 France, sergey.gavrilyuk@polytech.univ-mrs.fr}\ \
and\ \ H. GOUIN \thanks{
University of Aix-Marseille \& C.N.R.S. U.M.R. 6181, Case 322, Av.
Escadrille Normandie-Niemen, 13397 Marseille Cedex 20 France,
henri.gouin@univ-cezanne.fr}}
\maketitle

\begin{abstract}
The dynamics of the Reynolds stress tensor for turbulent flows is described
with an evolution equation coupling both geometric effects and turbulent
source terms. The effects of the mean flow geometry are shown up when the
source terms are neglected: the Reynolds stress tensor is then expressed as
the sum of three tensor products of vector fields which are governed by a
distorted gyroscopic equation. Along the mean flow trajectories, the
fluctuations of velocity are described by differential equations whose
coefficients depend only on the mean flow deformation. If the mean flow
vorticity is small enough, an approximate turbulence model is derived, and
its application to shear shallow water flows is proposed. Moreover, the
approximate turbulence model admits a variational formulation which is
similar to the one of capillary fluids.
\end{abstract}

\begin{epigraphs}
\qitem{Published in:\\"International Journal of Engineering Science\\ 
59 (2012) pp. 65-73"\\http://dx.doi.org/10.1016/j.ijengsci.2012.03.008}{}
\end{epigraphs}

\section{Introduction}

The system describing turbulent compressible barotropic flows is composed of
equation of mass balance, equation of average momentum and evolution
equation for the Reynolds stress tensor. In the following, we see that the
Reynolds stress tensor equation is mainly driven by the velocity gradient
tensor of the mean motion and this equation is the main object of our study
when the source term is negligible.

The reason for considering the simplified turbulence model without the
source term is twofold. First, in numerical studies of compressible
turbulent flows, this homogeneous equation is a natural step in applying the
splitting-up technique (see for example \cite{Berthon}). Secondly, such a
homogeneous system appears as an exact asymptotic model of weakly shearing
flows of long waves over a flat bottom (see \cite{Teshukov}). The only
difference is the space dimension: two dimensions are considered for shallow
water flows instead of three dimensions for the general case.

We use the spectral decomposition of the Reynolds stress tensor and in the
homogeneous case, we obtain a simpler dynamical system for the eigenvalues
and the eigenvectors of the Reynolds stress tensor. The system admits a
simple physical interpretation: the motion of each point of the turbulent
flow is analogous to the motion of a free rigid body moving along the mean
flow and rotating with an angular velocity which is different from the mean
flow vorticity. The angular velocity is completely determined by the mean
flow velocity. The moments of inertia of the free rigid body are not
constant, they are also determined by the mean flow.

When the mean flow vorticity is small enough, an approximate turbulence
model is obtained, which admits a variational formulation.

%Its application
%to the shear shallow water flows is considered.

\section{The governing equations}

The governing equations of barotropic turbulent compressible fluids are (see
\cite{Pironneau,Pope,Wilcox}) :
\begin{equation}
\left\{
\begin{array}{c}
\left\langle \rho \right\rangle _{t}+(\left\langle \rho \right\rangle
\,U_{i}),_{i}=0, \\
\\
\left( \left\langle \rho \right\rangle U_{i}\right) _{t}+\left( \left\langle
\rho \right\rangle \,U_{i}U_{j}+\left\langle p\right\rangle \,\delta
_{ij}+\left\langle \rho \,u_{i}u_{j}\right\rangle \right) ,_{j}=0, \\
\\
\left\langle \rho \,u_{i}u_{j}\right\rangle _{t}+\left( \left\langle \rho
\,u_{i}u_{j}\right\rangle \,U_{k}\right) ,_{k}+\left\langle \rho
\,u_{k}u_{j}\right\rangle \,U_{i,k}+\left\langle \rho
\,u_{i}u_{k}\right\rangle \,U_{j,k}=S_{ij}\,%
\end{array}%
\right.   \label{System1}
\end{equation}%
where \textquotedblleft brackets" mean the averaging, \textquotedblleft
coma" means the derivation with respect to the Eulerian coordinates $\mathbf{%
x}=\{x_{i}\},\ i\in \left\{ 1,2,3\right\} $ and index $"{t}"$ means the
partial derivative with respect to time, $\rho $ is the fluid density, $%
\mathbf{U=}\left\{ U_{i}\right\} $, $\,i\in \{1,2,3\}$ is the mass average
velocity, $p$ is the pressure, $\mathbf{u=\{}u_{i}\},\ i\in \left\{
1,2,3\right\} $ is the velocity fluctuation verifying $\left\langle \rho
\mathbf{u}\right\rangle =0$. Repeated indices mean summation. Here  $\mathbf{%
S}=\left\{ S_{ij}\right\} $ is a source term, and its explicit expression
can be written as \ :
\begin{equation*}
S_{ij}=-\left\langle \,u_{i}p,_{j}\right\rangle -\left\langle
\,u_{j}p,_{i}\right\rangle -\left\langle \rho \,u_{i}u_{j}u_{k}\right\rangle
,_{k}.
\end{equation*}
We introduce the Reynolds stress tensor
\begin{equation*}
\mathbf{R}=\left\langle \rho \ \mathbf{u\otimes u}\right\rangle \mathbf{,}%
\quad (R_{ij}=\left\langle \rho \,u_{i}u_{j}\right\rangle )\text{.}
\end{equation*}%
System (\ref{System1}) can be rewritten in the tensorial form
\begin{equation}
\left\{
\begin{array}{c}
\displaystyle\frac{\partial \left\langle \rho \right\rangle }{\partial t}+%
\mathrm{div}\left( \left\langle \rho \right\rangle \mathbf{U}\right) =0, \\
\\
\quad \displaystyle\left\langle \rho \right\rangle \frac{d\mathbf{U}}{dt}%
+\nabla \left\langle p\right\rangle +\left( \mathrm{div}\,\mathbf{R}\right)
^{T}=\mathbf{0}, \\
\\
\qquad \displaystyle\frac{d\mathbf{R}}{dt}+\mathbf{R}\,\mathrm{div}\mathbf{U}%
+\frac{\partial \mathbf{U}}{\partial \mathbf{x}}\,\mathbf{R}+\mathbf{R}%
\left( \frac{\partial \mathbf{U}}{\partial \mathbf{x}}\right) ^{T}=\mathbf{S}%
,%
\end{array}%
\right.   \label{System2}
\end{equation}%
where $\displaystyle d/dt\ $ means the material derivative with respect to
the mean motion
\begin{equation*}
\frac{d}{dt}=\frac{\partial }{\partial t}+\mathbf{U}^{T}\ \nabla \ .
\end{equation*}%
The superscript " ${T}\ $" \ means the transposition. Using the mass
conservation law, the equation for the volume Reynolds stress tensor $%
\mathbf{R}$ can be rewritten as the equation for the specific (or per unit
mass) Reynolds stress tensor
\begin{equation}
\frac{d\mathbf{P}}{dt}+\frac{\partial \mathbf{U}}{\partial \mathbf{x}}\
\mathbf{P}+\mathbf{P}\left( \frac{\partial \mathbf{U}}{\partial \mathbf{x}}%
\right) ^{T}=\frac{\mathbf{S}}{\left\langle \rho \right\rangle },
\label{Equation_3}
\end{equation}%
where%
\begin{equation*}
\mathbf{P}=\frac{\mathbf{R}}{\left\langle \rho \right\rangle }.
\end{equation*}%
The structure of source term  $\mathbf{S}$ has generated much debate within
physical and mathematical communities. Our goal is not to add new closure
hypotheses, but to study the structure of the "master" equation (\ref{Equation_3})
when $\mathbf{S=0.}$ The reason is twofold: \newline
- firstly, in the numerical study of compressible turbulent flows, this is a
natural step in applying the splitting-up technique (see for example \cite%
{Berthon}),\newline
- secondly, system (\ref{System2}) also appears as an exact asymptotic model
of weakly shearing flows of long waves (turbulent shallow water flows) over
a flat bottom (see \cite{Teshukov}) :
\begin{equation}
\left\{
\begin{array}{c}
\displaystyle\frac{\partial h}{\partial t}+\text{div}\left( h\,\mathbf{U}%
\right) =0, \\
\\
\displaystyle h\frac{d\mathbf{U}}{dt}+\nabla \left( \frac{gh^{2}}{2}\right)
+\left( \mathrm{{div}\,}\mathbf{R}\right) ^{T}=\mathbf{0}, \\
\\
\qquad \displaystyle\frac{d\mathbf{R}}{dt}+\mathbf{R}\,\mathrm{div}\mathbf{U}%
+\frac{\partial \mathbf{U}}{\partial \mathbf{x}}\ \mathbf{R}+\mathbf{R}%
\left( \frac{\partial \mathbf{U}}{\partial \mathbf{x}}\right) ^{T}=\mathbf{0}%
.%
\end{array}%
\right.   \label{Teshukov_system}
\end{equation}%
In system (\ref{Teshukov_system}), $h$ is the fluid depth playing the role
of the average density; the average pressure is given by $\left\langle
p\right\rangle =gh^{2}/2$, $g$ is the gravity acceleration, and
\begin{equation*}
\mathbf{R=}\int\limits_{0}^{h}\left[ (\widetilde{\mathbf{U}}-\mathbf{U}%
)\otimes (\widetilde{\mathbf{U}}-\mathbf{U})\right] dz,\qquad h\mathbf{U=}%
\int\limits_{0}^{h}\widetilde{\mathbf{U}}\,dz,
\end{equation*}%
where $\widetilde{\mathbf{U}}$ is the instantaneous velocity. Equations are
written for three-dimensional long waves and the production term is zero in
the limit of weakly shearing flows. Equations (\ref{Teshukov_system}) are
hyperbolic (see \cite{Teshukov} for proof). Finally, we will focus on the
equation of the Reynolds stress tensor per unit mass
\begin{equation}
\frac{d\mathbf{P}}{dt}+\frac{\partial \mathbf{U}}{\partial \mathbf{x}}\
\mathbf{P}+\mathbf{P}\left( \frac{\partial \mathbf{U}}{\partial \mathbf{x}}%
\right) ^{T}=0.  \label{basic_equation}
\end{equation}%
The case $\mathbf{S=0}$ corresponds to \emph{conservative motions of
turbulent compressible flows}; these motions verify the equations
\begin{equation}
\left\{
\begin{array}{c}
\displaystyle\frac{\partial \left\langle \rho \right\rangle }{\partial t}+%
\mathrm{div}\left( \left\langle \rho \right\rangle \mathbf{U}\right) =0, \\
\\
\quad \displaystyle\left\langle \rho \right\rangle \frac{d\mathbf{U}}{dt}%
+\nabla \left\langle p\right\rangle +\left( \mathrm{div}\,\mathbf{R}\right)
^{T}=\mathbf{0}, \\
\\
\qquad \displaystyle\frac{d\mathbf{R}}{dt}+\mathbf{R}\,\mathrm{div}\mathbf{U}%
+\frac{\partial \mathbf{U}}{\partial \mathbf{x}}\,\mathbf{R}+\mathbf{R}%
\left( \frac{\partial \mathbf{U}}{\partial \mathbf{x}}\right) ^{T}=\mathbf{0}%
.%
\end{array}%
\right.   \label{System6}
\end{equation}%
The particular case $\mathrm{{rot}\,}\mathbf{U}=0$ was investigated in \cite%
{Gouin}. In such a case $\left( \partial \mathbf{U}/\partial \mathbf{x}%
\right) ^{T}=\partial \mathbf{U}/\partial \mathbf{x}$ and Eq. (\ref%
{basic_equation}) corresponds to a two-covariant tensor convected by the
mean flow. This means that $\mathbf{P}$ has a zero Lie derivative $d_{L}\ $%
with respect to the velocity field $\mathbf{U}$ and the tensor $\mathbf{P}%
_{0}$, image of $\mathbf{P}$ in Lagrange coordinates $(t,\mathbf{X})$, only
depends on $\mathbf{X=}\left\{ X_{i}\right\} $, $\,i\in \{1,2,3\}$
\begin{equation}
d_{L}\mathbf{P}\equiv \frac{d\mathbf{P}}{dt}+\frac{\partial \mathbf{U}}{%
\partial \mathbf{x}}\ \mathbf{P}+\mathbf{P}\,\frac{\partial \mathbf{U}}{%
\partial \mathbf{x}}=0,\qquad \mathbf{P=}\left( F^{T}\right) ^{-1}\mathbf{P}%
_{0}\left( \mathbf{X}\right) F^{-1},  \label{Lie}
\end{equation}%
where $F=\partial \mathbf{x/}\partial \mathbf{X}$ is the deformation
gradient of the mean motion. \newline
The aim of the paper is the study of the homogeneous Reynolds stress tensor
equation structure (\ref{basic_equation}) in \textbf{\emph{the case}} $%
\mathtt{rot}\,\mathbf{U}\neq 0$.

\section{Geometric properties of the Reynolds stress tensor evolution}

The Reynolds stress tensor $\mathbf{P}$ is symmetric and semi-positive
definite. The tensor $\mathbf{P}$ can be rewritten in a local basis of
orthonormal eigenvectors in the form
\begin{equation*}
\mathbf{P}=\sum_{\alpha =1}^{3}\lambda _{\alpha }^{2}\ \mathbf{e}_{\alpha
}\otimes \mathbf{e}_{\alpha }\equiv \sum_{\alpha =1}^{3}\lambda _{\alpha
}^{2}\ \mathbf{e}_{\alpha }\,\mathbf{e}_{\alpha }^{T}.
\end{equation*}%
The eigenvalues $\lambda _{\alpha }^{2},$ $\alpha \in \{1,2,3\}$ are
non-negative; the case $\lambda _{\alpha }^{2}>0\ $is a generic one. For the
two-dimensional case, $\lambda _{3}^{2}\equiv 0.$ Let us denote
\begin{equation*}
\mathbf{a}_{\alpha }=\lambda _{\alpha }\ \mathbf{e}_{\alpha },\quad (\lambda
_{\alpha }>0)\qquad \alpha \in \{1,2,3\}.
\end{equation*}%
Then,%
\begin{equation}
\mathbf{P}=\sum_{\alpha =1}^{3}\mathbf{a}_{\alpha }\otimes \mathbf{a}%
_{\alpha }\equiv \sum_{\alpha =1}^{3}\mathbf{a}_{\alpha }\,\mathbf{a}%
_{\alpha }^{T}.  \label{decomposition2}
\end{equation}%
From Eq. (\ref{decomposition2}), we deduce
\begin{equation}
\frac{d\mathbf{P}}{dt}=\ \sum_{\alpha =1}^{3}\left[ \frac{d\mathbf{a}%
_{\alpha }}{dt}\ \mathbf{a}_{\alpha }^{T}+\mathbf{a}_{\alpha }\,\left( \frac{%
d\mathbf{a}_{\alpha }}{dt}\right) ^{T}\right] .  \label{reynolds derivative}
\end{equation}%
By using Eq. (\ref{reynolds derivative}), Eq. (\ref{basic_equation}) can be
written
\begin{equation}
\sum_{\alpha =1}^{3}\left( \frac{d\mathbf{a}_{\alpha }}{dt}+\frac{\partial
\mathbf{U}}{\partial \mathbf{x}}\ \mathbf{a}_{\alpha }\right) \mathbf{a}%
_{\alpha }^{T}+\left[ \left( \frac{d\mathbf{a}_{\alpha }}{dt}+\frac{\partial
\mathbf{U}}{\partial \mathbf{x}}\ \mathbf{a}_{\alpha }\right) \mathbf{a}%
_{\alpha }^{T}\right] ^{T}=0.  \label{reynoldsevolution derivative}
\end{equation}%
The vector $\displaystyle{d\mathbf{a}_{\alpha }}/{dt}+(\partial \mathbf{U}%
/\partial \mathbf{x})\,\mathbf{a}_{\alpha }$ can be developped in the local
basis $\{\mathbf{a}_{\beta }\},\ \beta \in \{1,2,3\}$ of eigenvectors; one
obtains
\begin{equation}
\frac{d\mathbf{a}_{\alpha }}{dt}+\frac{\partial \mathbf{U}}{\partial \mathbf{%
x}}\ \mathbf{a}_{\alpha }=\sum_{\beta =1}^{3}A_{\beta \alpha }\,\mathbf{a}%
_{\beta },\qquad \alpha \in \{1,2,3\}  \label{eigendecomposition}
\end{equation}%
where $A_{\beta \alpha },\ (\alpha ,\beta \in \{1,2,3\})$ are the scalar
components to be determined. By using Eq. (\ref{eigendecomposition}), Eq. (%
\ref{reynoldsevolution derivative}) leads to
\begin{equation*}
\sum_{\alpha =1}^{3}A_{\alpha \alpha }\,\mathbf{g}_{\alpha \alpha
}+\sum_{\alpha \neq \beta =1}^{3}\left( A_{\alpha \beta }+A_{\beta \alpha
}\right) \ \mathbf{g}_{\alpha \beta }=0,
\end{equation*}%
where $\mathbf{g}_{\alpha \alpha }=2\mathbf{a}_{\alpha }\,\mathbf{a}_{\alpha
}^{T}\ \ $and $\ \mathbf{g}_{\alpha \beta }=\mathbf{g}_{\beta \alpha }=%
\mathbf{a}_{\alpha }\,\mathbf{a}_{\beta }^{T}+\mathbf{a}_{\beta }\,\mathbf{a}%
_{\alpha }^{T},\ (\alpha ,\beta \in \{1,2,3\})$, are six independent
symmetric tensors. Consequently,
\begin{equation*}
A_{\alpha \alpha }=0\text{ \ \ and \ \ }A_{\alpha \beta }+A_{\beta \alpha
}=0,\qquad \alpha ,\beta \in \{1,2,3\}.
\end{equation*}%
Equation (\ref{reynoldsevolution derivative}) is equivalent to
\begin{equation}
\frac{d\mathbf{a}_{\alpha }}{dt}+\frac{\partial \mathbf{U}}{\partial \mathbf{%
x}}\ \mathbf{a}_{\alpha }=\Lambda \circ i\left( \mathbf{\pi }\right) \mathbf{%
e}_{\alpha }\ \ \text{with\ \ }\mathbf{\pi =}A_{32\ }\mathbf{e}_{1}+A_{13\ }%
\mathbf{e}_{2}+A_{21\ }\mathbf{e}_{3},\quad \alpha \in \{1,2,3\},
\label{aevolution}
\end{equation}%
where a diagonal matrix $\Lambda $ and an antisymmetric matrix $i\left(
\mathbf{\pi }\right) $ are determined in the basis $\mathbf{e}_{\beta },\
\beta \in \{1,2,3\}$ as
\begin{equation*}
\Lambda =\left(
\begin{array}{ccc}
\lambda _{1} & 0 & 0 \\
0 & \lambda _{2} & 0 \\
0 & 0 & \lambda _{3}%
\end{array}%
\right) ,\qquad i\left( \mathbf{\pi }\right) =\left(
\begin{array}{ccc}
0 & -A_{21\ } & A_{13} \\
\ \ A_{21\ } & 0 & -A_{32\ } \\
-A_{13} & A_{32\ } & 0%
\end{array}%
\right) .
\end{equation*}%
The vectors $\mathbf{a}_{\beta },\ \beta \in \{1,2,3\}$ are orthogonal, $%
\mathbf{a}_{\alpha }^{T}\ \mathbf{a}_{\beta }=0,\ (\alpha \neq \beta )$.
This is equivalent to
\begin{equation}
\mathbf{a}_{\alpha }^{T}\ \frac{d\mathbf{a}_{\beta }}{dt}+\mathbf{a}_{\beta
}^{T}\ \frac{d\mathbf{a}_{\alpha }}{dt}=0.  \label{orthogonality}
\end{equation}%
Consequently, Eqs. (\ref{aevolution}) - (\ref{orthogonality}) yield

\begin{equation*}
\begin{array}{l}
\forall \,\alpha \neq \beta \in \{1,2,3\}, \\
\\
\displaystyle\mathbf{a}_{\alpha }^{T}\ \left( \Lambda \circ i\left( \mathbf{%
\pi }\right) \mathbf{e}_{\beta }-\frac{\partial \mathbf{U}}{\partial \mathbf{%
x}}\ \mathbf{a}_{\beta }\right) +\mathbf{a}_{\beta }^{T}\ \left( \Lambda
\circ i\left(\mathbf{\pi }\right) \mathbf{e}_{\alpha }-\frac{\partial
\mathbf{U}}{\partial \mathbf{x}}\ \mathbf{a}_{\alpha }\right) =0.%
\end{array}%
\end{equation*}%
\newline

Or%
\begin{equation}
\begin{array}{r}
2\lambda _{\alpha }\lambda _{\beta }\mathbf{e}_{\alpha }^{T}\ \mathbf{D}\
\mathbf{e}_{\beta }=\mathbf{a}_{\alpha }^{T}\ \Lambda \circ i\left( \mathbf{%
\pi }\right) \mathbf{e}_{\beta }+\mathbf{a}_{\beta }^{T}\ \Lambda \circ
i\left( \mathbf{\pi }\right) \mathbf{e}_{\alpha }\  \\
\\
\equiv \lambda _{\alpha }^{2}\mathbf{e}_{\alpha }^{T}\ i\left( \mathbf{\pi }%
\right) \mathbf{e}_{\beta }+\lambda _{\beta }^{2}\mathbf{e}_{\beta }^{T}\
i\left( \mathbf{\pi }\right) \mathbf{e}_{\alpha }%
\end{array}
\label{Deltavectors}
\end{equation}%
where
\begin{equation*}
\mathbf{D}=\displaystyle\frac{1}{2}\,\left( \frac{\partial \mathbf{U}}{%
\partial \mathbf{x}}+\left( \frac{\partial \mathbf{U}}{\partial \mathbf{x}}%
\right) ^{T}\right) \,
\end{equation*}%
is the rate of deformation tensor corresponding to the mean flow. We denote
the mixed product of three vectors\ $\left\{ \mathbf{a,b,c}\right\} $ as $%
\left( \mathbf{a,b,c}\right) \equiv \mathbf{a}^{T}\left( \mathbf{b\wedge c}%
\right) $. Hence, Eq. (\ref{Deltavectors}) yields

\begin{equation*}
\begin{array}{r}
2\lambda _{\alpha }\lambda _{\beta }\,\mathbf{e}_{\alpha }^{T}\ \mathbf{D}\
\mathbf{e}_{\beta }=\left( \ \lambda _{\alpha }^{2}-\lambda _{\beta
}^{2}\right) \left( \mathbf{e}_{\alpha },\mathbf{\pi ,e}_{\beta }\right) \\
\\
=\left( \ \lambda _{\beta }^{2}-\lambda _{\alpha }^{2}\right) \left( \mathbf{%
\pi ,e}_{\alpha },\mathbf{e}_{\beta }\right) \\
\\
=\left( \ \lambda _{\beta }^{2}-\lambda _{\alpha }^{2}\right) \mathbf{\pi }%
^{T}\mathbf{e}_{\gamma },%
\end{array}%
\end{equation*}%
where $\left\{ \alpha ,\beta ,\gamma \right\} $ is a cyclic permutation of
the triplet $\left\{ 1,2,3\right\} $. Finally, we get
\begin{equation}
\mathbf{\pi }^{T}\mathbf{e}_{\gamma }=\frac{2\lambda _{\alpha }\lambda
_{\beta }\ \mathbf{e}_{\alpha }^{T}\ \mathbf{D}\ \mathbf{e}_{\beta }}{\
\lambda _{\beta }^{2}-\lambda _{\alpha }^{2}}.  \label{pi-formula}
\end{equation}%
Equation (\ref{aevolution}) can be written
\begin{equation}
\lambda _{\alpha }\frac{d\mathbf{e}_{\alpha }}{dt}+\frac{d\lambda _{\alpha }%
}{dt}\,{\mathbf{e}}_{\alpha }+\lambda _{\alpha }\frac{\partial \mathbf{U}}{%
\partial \mathbf{x}}\,\mathbf{e}_{\alpha }\ -\Lambda \circ i\left( \mathbf{%
\pi }\right) \mathbf{e}_{\alpha }=0,\qquad \alpha \in \{1,2,3\}.
\label{equadiff1}
\end{equation}%
Since
\begin{equation*}
{\mathbf{e}}_{\alpha }^{T}\frac{d\mathbf{e}_{\alpha }}{dt}=0,\qquad {\mathbf{%
e}}_{\alpha }^{T}\Lambda \circ i\left( \mathbf{\pi }\right) \mathbf{e}%
_{\alpha }=\lambda _{\alpha }{\mathbf{e}}_{\alpha }^{T}i\left( \mathbf{\pi }%
\right) \mathbf{e}_{\alpha }=0,
\end{equation*}%
by multiplying the left side of Eq. (\ref{equadiff1}) with ${\mathbf{e}}%
_{\alpha }^{T}$, we get
\begin{equation*}
\frac{d\lambda _{\alpha }}{dt}+\ \left( \mathbf{e}_{\alpha }^{T}\ \mathbf{D}%
\ \mathbf{e}_{\alpha }\right) \lambda _{\alpha }=0.
\end{equation*}%
By multiplying the left side of Eq. (\ref{equadiff1}) with the projector $%
\left( \mathbf{I}-\mathbf{e}_{\alpha }\mathbf{e}_{\alpha }^{T}\right) $, we
get
\begin{equation*}
\lambda _{\alpha }\frac{d\mathbf{e}_{\alpha }}{dt}+\lambda _{\alpha }\left(
\mathbf{I}-\mathbf{e}_{\alpha }\mathbf{e}_{\alpha }^{T}\right) \frac{%
\partial \mathbf{U}}{\partial \mathbf{x}}\,\mathbf{e}_{\alpha }\ -\Lambda
\circ i\left( \mathbf{\pi }\right) \mathbf{e}_{\alpha }=0.
\end{equation*}%
We will prove that there exists a vector $\mathbf{\Pi }$ such that
\begin{equation*}
\frac{d\mathbf{e}_{\alpha }}{dt}\ =\mathbf{\Pi }\wedge \mathbf{e}_{\alpha },
\end{equation*}%
Such a vector $\mathbf{\Pi }$ should verify the condition
\begin{equation}
\lambda _{\alpha }i\left( \mathbf{\Pi }\right) \mathbf{e}_{\alpha }+\lambda
_{\alpha }\left( \mathbf{I}-\mathbf{e}_{\alpha }\mathbf{e}_{\alpha
}^{T}\right) \frac{\partial \mathbf{U}}{\partial \mathbf{x}}\,\mathbf{e}%
_{\alpha }\ -\Lambda \circ i\left( \mathbf{\pi }\right) \mathbf{e}_{\alpha
}=0.  \label{condition}
\end{equation}%
By multiplying Eq. (\ref{condition}) with $\mathbf{e}_{\beta }^{T}$ where $%
\beta \neq \alpha ,$ we get
\begin{equation}
\lambda _{\alpha }\mathbf{\Pi }^{T}\mathbf{e}_{\gamma }=\lambda _{\beta }%
\mathbf{\pi }^{T}\mathbf{e}_{\gamma }-\lambda _{\alpha }\mathbf{e}_{\beta
}^{T}\frac{\partial \mathbf{U}}{\partial \mathbf{x}}\,\mathbf{e}_{\alpha }.
\label{condition2}
\end{equation}%
By replacing Rel. (\ref{pi-formula}) into Eq. (\ref{condition2}) we get

\begin{equation*}
\begin{array}{r}
\displaystyle\mathbf{\Pi }^{T}\mathbf{e}_{\gamma }=\displaystyle\frac{%
\lambda _{\beta }^{2}\mathbf{e}_{\alpha }^{T}\left( \displaystyle\frac{%
\partial \mathbf{U}}{\partial \mathbf{x}}+\left( \frac{\partial \mathbf{U}}{%
\partial \mathbf{x}}\right) ^{T}\right) \mathbf{e}_{\beta }}{\ \lambda
_{\beta }^{2}-\lambda _{\alpha }^{2}}-\mathbf{e}_{\beta }^{T}\frac{\partial
\mathbf{U}}{\partial \mathbf{x}}\,\mathbf{e}_{\alpha } \\
\\
\displaystyle=\frac{\mathbf{e}_{\beta }^{T}\left( \lambda _{\beta
}^{2}\left( \displaystyle\frac{\partial \mathbf{U}}{\partial \mathbf{x}}%
\right) ^{T}+\lambda _{\alpha }^{2}\displaystyle\frac{\partial \mathbf{U}}{%
\partial \mathbf{x}}\right) \mathbf{e}_{\alpha }\,}{\ \lambda _{\beta
}^{2}-\lambda _{\alpha }^{2}}\, .%
\end{array}%
\end{equation*}%
\newline
Now, we can formulate the result, \newline

\textbf{Theorem 1. }\emph{The Reynolds stress tensor can be written in the
form }
\begin{equation*}
\mathbf{R}=\left\langle \rho \right\rangle \sum_{\alpha =1}^{3}\lambda
_{\alpha }^{2}\ \mathbf{e}_{\alpha }\otimes \mathbf{e}_{\alpha }.
\end{equation*}%
\emph{\ The eigenvectors }$\mathbf{e}_{\alpha }$ \emph{and the eigenvalues }$%
\lambda _{\alpha }$ \emph{verify the equations: }
\begin{equation}
\ \ \ \ \ \ \ \ \ \ \ \ \ \ \ \ \ \ \ \left\{
\begin{array}{l}
\displaystyle\frac{d{\mathbf{e}}_{\alpha }}{dt}\ =\mathbf{\Pi }\wedge {%
\mathbf{e}}_{\alpha }, \\
\qquad \qquad \qquad \qquad \qquad \qquad \qquad \qquad \ \alpha \in
\{1,2,3\} \\
\displaystyle\frac{d\left( \text{Ln}\,\lambda _{\alpha }^{2}\right) }{dt}%
=-2\ \mu _{\alpha },%
\end{array}%
\right.  \label{driving system}
\end{equation}%
\newline

\emph{where}%
\begin{equation*}
\mu _{\alpha }=\mathbf{e}_{\alpha }^{T}\ \mathbf{D}\ \mathbf{e}_{\alpha
},\qquad \mathbf{\Pi }^{T}\mathbf{e}_{\gamma }=\frac{\mathbf{e}_{\beta
}^{T}\left( \lambda _{\beta }^{2}\left( \displaystyle\frac{\partial \mathbf{U%
}}{\partial \mathbf{x}}\right) ^{T}+\lambda _{\alpha }^{2}\displaystyle\frac{%
\partial \mathbf{U}}{\partial \mathbf{x}}\right) \mathbf{e}_{\alpha }\,}{\
\lambda _{\beta }^{2}-\lambda _{\alpha }^{2}}.
\end{equation*}%
\emph{The triplet }$\left\{ \alpha ,\beta ,\gamma \right\} \ $\emph{%
corresponds to a cyclic permutation of \ the triplet }$\left\{ 1,2,3\right\}
.$\emph{\ } \newline

Equation (\ref{driving system}$_{1}$) is similar to the equations of a rigid
body (see \cite{Marsden}). The vectors ${\mathbf{e}}_{\alpha }$ form a
natural moving frame $\left\{ \mathbf{e}_{\alpha }\right\} _{\alpha =1}^{3}$
whose evolution is determined by the mean rate of deformation tensor. The
eigenvalues $\lambda _{\alpha }^{2}$ of the Reynolds stress tensor are
determined by the evolution equation (\ref{driving system}$_{2}$). Let us
note that, if $\lambda _{\alpha }$ are initially positive, they will be
positive for any time. Hence, it means that the tensor $P$ will always be
positive definite. Due to the mass conservation law (\ref{System2}$_{1}$)\
and Eq. (\ref{driving system}$_{2}$), we obtain the following quantity
conserved along trajectories of mean flow:
\begin{equation*}
\frac{d}{dt}\left( \left\langle \rho \right\rangle ^{-2}\prod\limits_{\alpha
=1}^{3}\lambda _{\alpha }^{2}\right) =0.
\end{equation*}%
Consequently, system (\ref{driving system}) admits an invariant scalar along
the trajectories of mean flow. This invariant was earlier obtained in \cite%
{Debieve,Gouin} in a different form. Let us introduce the turbulent specific
energy
\begin{equation*}
e_{T}=\frac{1}{2}\,\mathtt{tr}\,\mathbf{P}=\frac{1}{2}\sum\limits_{\alpha
=1}^{3}\lambda _{\alpha }^{2}.
\end{equation*}

In the incompressible (isochoric) case, we have ${d\langle \rho \rangle }/{dt%
}=0$; the turbulent energy is minimal in the isotropic case when the three
eigenvalues $\lambda _{\alpha }^{2}$ are equal ($\lambda _{1}^{2}=\lambda
_{2}^{2}=\lambda _{3}^{2}=\lambda ^{2}$). In this case, the orthonormal
eigenvectors $\mathbf{e}_{\alpha },\ \alpha \in \{1,2,3\}$ of the Reynolds
stress tensor $\mathbf{P}$ are also the orthonormal eigenvectors of the mean
rate of deformation tensor $\mathbf{D}$ and $\mu _{\alpha }$ are the
corresponding eigenvalues.

In the compressible isotropic case $e_{T}=\left\langle \rho \right\rangle
^{2/3}\kappa $, $\kappa =3\lambda ^{2}/(2\left\langle \rho \right\rangle
^{2/3})$, and $\kappa $ is a classical invariant of isotropic compressible
turbulence.

In presence of shock waves the quantity $\left\langle \rho \right\rangle
^{-2}\prod\limits_{\alpha =1}^{3}\lambda _{\alpha }^{2}$ is not conserved
through shocks; it increases like the classical entropy in compressible
fluid dynamics. The estimation of the jump of turbulence entropy in
isotropic case was given in \cite{Gavrilyuk}.

As a consequence, the governing equations (\ref{System6}) admit the energy
conservation law
\begin{equation*}
\frac{\partial }{\partial t}\left( {\langle \rho \rangle }\left( \frac{1}{2}{%
\left\vert \mathbf{U}\right\vert ^{2}}+e_{i}+e_{T}\right) \right) + \mathrm{%
div} \left( {\langle \rho \rangle }\mathbf{U}\left( \frac{1}{2}{\left\vert
\mathbf{U}\right\vert ^{2}}+e_{i}+e_{T}\right) +\left( {\langle p\rangle
\mathbf{I}}+\mathbf{R}\right) \mathbf{U}\right) =0
\end{equation*}%
where the internal specific energy $e_{i}$ is defined by
\begin{equation*}
de_{i}=-{\langle p\rangle\, d}\left( \frac{1}{{\langle \rho \rangle }}%
\right) \
\end{equation*}%
and the mean pressure ${\langle p\rangle }$ is supposed to be a given
function of ${\langle \rho \rangle }$. Indeed, using (\ref{driving system}$%
_{2}$) we immediately obtain
\begin{equation*}
\frac{\partial }{\partial t}\left( {\langle \rho \rangle }\left( \frac{1}{2}{%
\left\vert \mathbf{U}\right\vert ^{2}}+e_{i}+e_{T}\right) \right) +\mathbf{%
\mathrm{div}}\left( {\langle \rho \rangle }\mathbf{U}\left( \frac{1}{2}{%
\left\vert \mathbf{U}\right\vert ^{2}}+e_{i}+e_{T}\right) +\left( \langle
p\rangle \mathbf{I}+\mathbf{R}\right) \mathbf{U}\right)
\end{equation*}%
\begin{equation*}
={\langle \rho \rangle }\frac{de_{T}}{dt}+\mathtt{tr}\left( \mathbf{R\,D}%
\right) =\frac{{\langle \rho \rangle }}{2}\frac{d}{dt}\left(
\sum\limits_{\alpha =1}^{3}\lambda _{\alpha }^{2}\right) +{\langle \rho
\rangle }\,\mathtt{tr}\left( \sum\limits_{\alpha =1}^{3}\lambda _{\alpha
}^{2}\mu _{\alpha }\right) =0.
\end{equation*}%
System (\ref{System6}) is a conservative and Galilean invariant system of
equations which is the counterpart of the Euler equations for the turbulent
compressible flows. The equations for the Reynolds stress tensor (\ref%
{System6}$_{3} $) are rewritten in a simpler form admitting a clear physical
interpretation.

\section{An approximate model}

In this Section, we derive a useful approximation of model (\ref{System6})
describing compressible turbulent flows for motions characterized by a small
average vorticity.

\subsection{Preliminaries}

Equation ({\ref{basic_equation}) can be rewritten as :
\begin{equation}
\frac{d\mathbf{P}}{dt}+\left( \frac{\partial \mathbf{U}}{\partial \mathbf{x}}%
\right) ^{T}\mathbf{P}+\mathbf{P}\ \frac{\partial \mathbf{U}}{\partial
\mathbf{x}}=\left[ \text{\ }i\left( \mathrm{rot}\mathbf{U}\right) \mathbf{\
,\ P}\,\right]  \label{Equation_speciale}
\end{equation}%
}with$\ $%
\begin{equation*}
\,\left[ \text{\ }i\left( \mathrm{rot}\mathbf{U}\right) \mathbf{\ ,\ P}\,%
\right] =\mathbf{P}\,i\left( \mathrm{rot}\mathbf{U}\right) -i\left( \mathrm{%
rot}\mathbf{U}\right) \mathbf{P},\ \ \ i\left( \mathrm{rot}\mathbf{U}\right)
=\frac{\partial \mathbf{U}}{\partial \mathbf{x}}\,-\left( \frac{\partial
\mathbf{U}}{\partial \mathbf{x}}\right) ^{T}.
\end{equation*}
Let $\tau $ be a characteristic time scale and $\omega $ a characteristic
value of the mean vorticity norm $\left\Vert \mathrm{rot}\mathbf{U}%
\right\Vert $; we assume that
\begin{equation}
\tau \,\omega \ll 1.  \label{order}
\end{equation}%
Relation (\ref{order}) is verified, in particular, for motions which are
close to one-dimensional ones. Equation (\ref{Equation_speciale}) gets the
form
\begin{equation}
\frac{d\mathbf{P}}{dt}+\left( \frac{\partial \mathbf{U}}{\partial \mathbf{x}}%
\right) ^{T}\mathbf{P}+\mathbf{P}\ \frac{\partial \mathbf{U}}{\partial
\mathbf{x}}=0.  \label{EqSpecialReynolds}
\end{equation}%
Using the solution (\ref{Lie}) when $\mathbf{P}_{0}(\mathbf{X})=\mathbf{I}$,
we consider the Reynolds stress tensor in the form
\begin{equation*}
\mathbf{P=}\sum_{\alpha }\nabla \ \varphi _{\alpha }\otimes \nabla \ \varphi
_{\alpha },
\end{equation*}%
where index $\alpha $ ranges over a finite number of integers and $\varphi
_{\alpha }$ are generalized Lagrangian coordinates :
\begin{equation*}
\frac{d\varphi _{\alpha }}{dt}=0.
\end{equation*}%
Covectors
\begin{equation*}
\mathbf{b}_{\alpha }^{T}=\frac{\partial \varphi _{\alpha }}{\partial \mathbf{%
x}}
\end{equation*}%
verify the identity

\begin{equation*}
\frac{d\mathbf{b}_{\alpha }^{T}}{dt}+\mathbf{b}_{\alpha }^{T}\frac{\partial
\mathbf{U}}{\partial \mathbf{x}}=\mathbf{0},
\end{equation*}%
corresponding to a zero Lie derivative of $\mathbf{b}_{\alpha }^{T}$ with
respect to the mass average velocity. In such a case, Eq. (\ref%
{EqSpecialReynolds}) is identically verified. Symmetric tensor $\mathbf{P}$
is determined by six scalar fields $\varphi _{\alpha },\ \alpha \in \left\{
1,\cdots ,6\right\} $. If, initially, vectors $\nabla \ \varphi _{\alpha },\
\alpha \in \left\{ 1,2,3\right\} $ constitute an orthogonal system
corresponding to the eigenvectors of $\mathbf{P}$, we can choose $\varphi
_{\alpha }=0,\, \alpha \in \left\{ 4,5,6\right\} $ and consequently,

\begin{equation}
\mathbf{P=}\sum_{\alpha =1}^{3}\nabla \ \varphi _{\alpha }\otimes \nabla \
\varphi _{\alpha }.  \label{special_Reynold_tensor}
\end{equation}%
It is worth to note that since the Reynolds stress tensor is positive
definite, it can be considered as a metric tensor of a Riemannian space
associated with the metric
\begin{equation*}
P_{ij}=\sum_{\alpha =1}^{3}\frac{\partial \varphi _{\alpha }}{\partial x_{i}}%
\frac{\partial \varphi _{\alpha }}{\partial x_{j}}.
\end{equation*}

This metric is flat because%
\begin{equation*}
\sum_{i,j}P_{ij}dx^{i}dx^{j}=\sum_{\alpha =1}^{3}\left( d\varphi _{\alpha
}\right) ^{2}
\end{equation*}

It is interesting to note that this special structure of the Reynolds stress
tensor implies a variational structure of equations (\ref{System6}).

\subsection{The Hamilton principle}

The aim of this Section is to prove that in special form (\ref%
{special_Reynold_tensor}), System (\ref{System6}) admits a variational
formulation which is similar to the one of capillary fluids \cite%
{Cahn,Casal,Casal_2,Eglit,Dell'Isola,Truskinovsky,van_der_waals}. However,
in our case, the expression of the \emph{capillary energy} is determined by
the gradients of three scalar order parameters transported along the
trajectories of the mean flow. \newline
We consider the specific internal energy in the form
\begin{equation*}
e_{i}=e_{i}\left( \left\langle \rho \right\rangle \right) .
\end{equation*}%
The mean density is submitted to the constraint
\begin{equation*}
\frac{\partial \left\langle \rho \right\rangle }{\partial t}+\mathrm{div}%
\left( \left\langle \rho \right\rangle \mathbf{U}\right) =0.
\end{equation*}%
Let us define the specific turbulent energy as
\begin{equation*}
e_{T}=\sum\limits_{\alpha =1}^{3}\frac{\left\vert \nabla \varphi _{\alpha
}\right\vert ^{2}}{2},  \label{turbulent energy}
\end{equation*}%
where scalars $\varphi _{\alpha }$ are submitted to the constraint
\begin{equation*}
\frac{d\varphi _{\alpha }}{dt}=0,\qquad \alpha \in {1,2,3}.
\end{equation*}%
For a material volume $D_{t}$ of the mean motion, the Hamilton action
calculated between times $t_{1}$, $t_{2}$ is
\begin{equation*}
a=\int_{t_{1}}^{t_{2}}\left( \iiint_{D_{t}}{\langle \rho \rangle }\mathcal{%
L\ }dv\right) dt,
\end{equation*}%
with the specific Lagrangian
\begin{equation*}
\mathcal{L}=\left( \frac{1}{2}\,\mathbf{U}^{T}\mathbf{U}-e_{i}-e_{T}\right) .
\end{equation*}%
The fluid motion is a $C^{2}$- diffeomorphisme $\phi $ from a
three-dimensional space $D_{0}$ into the physical space $D_{t}$ :
\begin{equation*}
\mathbf{x}=\phi \left( \mathbf{X},t\right) \ \ \ \mathrm{or}\ \ \ x_{i}=\phi
_{i}(X_{1},X_{2},X_{3},t)\,,\quad i\in \{1,2,3\}.
\end{equation*}
Let a one-parameter family of virtual motions denoted by $\{\phi
_{\varepsilon }\}$, possessing continuous derivatives up to the second order
and expressed in the form :
\begin{equation*}
\mathbf{x=\Phi }\left( \mathbf{X},t;\varepsilon \right) ,
\end{equation*}%
with $\varepsilon \in \mathcal{O},$ where $\mathcal{O}$ is an open interval
containing $0$ and such that $\mathbf{\Phi }\left( \mathbf{X},t;0\right)
=\phi \left( \mathbf{X,}t\right) $ (the real motion of the continuous medium
is obtained when $\varepsilon =0$). The derivation with respect to $%
\varepsilon $ when $\varepsilon =0$ is denoted by $\delta $. Derivation $%
\delta $ is named variation and the virtual displacement $\delta {\phi }$ is
the variation of the motion of the medium. At time $t$, the virtual
displacement of the particle $\mathbf{x}$ is $\delta \mathbf{x}\,$ obtained
when $\delta \mathbf{X}=0$ and $\delta \varepsilon =1$ at $\varepsilon =0$;
the virtual displacement corresponds to the field of tangent vectors to $%
D_{t}$
\begin{equation*}
{\mathbf{x}}\in D_{t}\ \mathbf{\rightarrow \mathbf{\zeta }}=\mathbf{\psi (x)}%
\equiv \frac{\partial \mathbf{\Phi }}{\partial \varepsilon }\left\vert
_{\varepsilon =0}\right. \in T_{\mathbf{x}}(D_{t})\,,
\end{equation*}%
where $T_{\mathbf{x}}(D_{t})$ is the tangent vector bundle to $D_{t}$ at $%
\mathbf{x}$. \newline
The Hamilton principle reads : \textit{for each vector field of virtual
displacements such that\ }$\mathbf{\mathbf{\zeta \ }}$\textit{and its
derivatives vanish at} \textit{the boundary}\ $\partial \Omega $ \textit{of}
$\Omega $,
\begin{equation*}
\delta a=\ 0.
\end{equation*}

We have the following general results (see \cite%
{Berdichevsky,Casal,Gavrilyuk2,Serrin}) :
\begin{equation*}
\left\{
\begin{array}{l}
\displaystyle\delta \left( \int_{t_{1}}^{t_{2}}\left( \iiint_{D_{t}}{\langle
\rho \rangle }\,\mathcal{L\ }dv\right) dt\right) =\int_{\Omega }{\langle
\rho \rangle \,\delta }\mathcal{L\ }dv\,dt, \\
\\
\displaystyle\delta \mathbf{U=}\frac{d\mathbf{\mathbf{\zeta }}}{dt}, \\
\\
\displaystyle\delta \left\langle \rho \right\rangle =-\left\langle \rho
\right\rangle \mathrm{{div}\mathbf{\mathbf{\zeta }},} \\
\\
\displaystyle\delta \left( \frac{\partial \varphi _{\alpha }}{\partial
\mathbf{x}}\right) =-\frac{\partial \varphi _{\alpha }}{\partial \mathbf{x}}%
\frac{\partial \mathbf{\mathbf{\zeta }}}{\partial \mathbf{x}},%
\end{array}%
\right.
\end{equation*}%
where $\Omega =\left[ t_{1},t_{2}\right] \times D_{t}$ and $\int_{\Omega }$
is the quadruple integral in the time-space domain $\Omega $. Consequently,
\begin{equation*}
{\langle \rho \rangle \,\delta }\mathcal{L}={\langle \rho \rangle \,}\mathbf{%
U}^{T}\frac{d\mathbf{\mathbf{\zeta }}}{dt}+\left\langle \rho \right\rangle
^{2}\frac{\partial e_{i}}{\partial {\langle \rho \rangle }}\;\mathrm{{div}%
\mathbf{\mathbf{\zeta +}}\sum\limits_{\alpha =1}^{3}{\langle \rho \rangle \,}%
\frac{\partial \varphi _{\alpha }}{\partial \mathbf{x}}\frac{\partial
\mathbf{\mathbf{\zeta }}}{\partial \mathbf{x}}\left( \frac{\partial \varphi
_{\alpha }}{\partial \mathbf{x}}\right) ^{T}.}
\end{equation*}%
Let us denote $\left\langle p\right\rangle =\left\langle \rho \right\rangle
^{2}\partial e_{i}/\partial {\langle \rho \rangle }$, the mean pressure
scalar field of the fluid; due to the identities,
\begin{equation*}
\left\{
\begin{array}{l}
\displaystyle{\langle \rho \rangle \,}\mathbf{U}^{T}\frac{d\mathbf{\mathbf{%
\zeta }}}{dt}\equiv \frac{\partial \left( {\langle \rho \rangle \,}\mathbf{U}%
^{T}\mathbf{\mathbf{\zeta }}\right) }{\partial t}+\mathrm{{div}\left( {%
\langle \rho \rangle \,}\mathbf{U}^{T}\mathbf{\mathbf{\zeta }}\right) -{%
\langle \rho \rangle }\frac{d\mathbf{U}^{T}}{dt}\,\mathbf{\mathbf{\zeta }},}
\\
\\
\displaystyle\left\langle p\right\rangle \mathrm{{div}\mathbf{\mathbf{\zeta }%
} \equiv {div}\left( \left\langle p\right\rangle \,\mathbf{\mathbf{\zeta }}%
\right) -\frac{\partial \left\langle p\right\rangle }{\partial \mathbf{x}}\,%
\mathbf{\mathbf{\zeta }},} \\
\\
\displaystyle\sum\limits_{\alpha =1}^{3}{\langle \rho \rangle \,}\frac{%
\partial \varphi _{\alpha }}{\partial \mathbf{x}}\frac{\partial \mathbf{%
\mathbf{\zeta }}}{\partial \mathbf{x}}\left( \frac{\partial \varphi _{\alpha
}}{\partial \mathbf{x}}\right) ^{T}\equiv \sum\limits_{\alpha =1}^{3}\text{tr%
}\left( {\langle \rho \rangle \,}\left( \frac{\partial \varphi _{\alpha }}{%
\partial \mathbf{x}}\right) ^{T}\frac{\partial \varphi _{\alpha }}{\partial
\mathbf{x}}\frac{\partial \mathbf{\mathbf{\zeta }}}{\partial \mathbf{x}}%
\right) , \\
\displaystyle\equiv \sum\limits_{\alpha =1}^{3}\text{div}\left( {\langle
\rho \rangle \,}\left( \frac{\partial \varphi _{\alpha }}{\partial \mathbf{x}%
}\right) ^{T}\frac{\partial \varphi _{\alpha }}{\partial \mathbf{x}}\,%
\mathbf{\mathbf{\zeta }}\right) -\sum\limits_{\alpha =1}^{3}\text{div}\left(
{\langle \rho \rangle \,}\left( \frac{\partial \varphi _{\alpha }}{\partial
\mathbf{x}}\right) ^{T}\frac{\partial \varphi _{\alpha }}{\partial \mathbf{x}%
}\right) \,\mathbf{\mathbf{\zeta }}\,. \\
\end{array}%
\right.
\end{equation*}%
Stokes' formula implies that terms $\displaystyle\frac{\partial \left( {%
\langle \rho \rangle \,}\mathbf{U}^{T}\mathbf{\mathbf{\zeta }}\right) }{%
\partial t},\ \mathrm{{div}\left( {\langle \rho \rangle \,}\mathbf{U}^{T}%
\mathbf{\mathbf{\zeta }}\right) ,\ {div}\left( \left\langle p\right\rangle \,%
\mathbf{\mathbf{\zeta }}\right) }$ and\newline
$\ \mathrm{div}\displaystyle\left( {\langle \rho \rangle \,}\left( \frac{%
\partial \varphi _{\alpha }}{\partial \mathbf{x}}\right) ^{T}\frac{\partial
\varphi _{\alpha }}{\partial \mathbf{x}}\,\mathbf{\mathbf{\zeta }}\right) $
can be integrated on $\partial \Omega $ where $\mathbf{\mathbf{\zeta }}$
vanishes. We get for each field of virtual motion,
\begin{equation*}
\delta a=-\int_{\Omega }\left\{ {\langle \rho \rangle }\frac{d\mathbf{U}^{T}%
}{dt}+\frac{\partial \left\langle p\right\rangle }{\partial \mathbf{x}}%
+\sum\limits_{\alpha =1}^{3}\text{div}\left( {\langle \rho \rangle \,}\left(
\frac{\partial \varphi _{\alpha }}{\partial \mathbf{x}}\right) ^{T}\frac{%
\partial \varphi _{\alpha }}{\partial \mathbf{x}}\right) \right\} \mathbf{%
\mathbf{\zeta }}\mathcal{\ }dv\,dt,
\end{equation*}%
and the fundamental lemma of variational calculus yields the equation of
motion
\begin{equation*}
\left\langle \rho \right\rangle \frac{d\mathbf{U}}{dt}+\nabla \left\langle
p\right\rangle +\mathrm{div}\,\left( \left\langle \rho \right\rangle
\sum\limits_{\alpha =1}^{3}\nabla \varphi _{\alpha }\otimes \nabla \varphi
_{\alpha }\right) =0;
\end{equation*}%
the variational formulation of the approximate system is established.

%\subsection{Shear shallow water flows}

%We present the approximate model in the case of shear shallow water flows \
%described by (\ref{Teshukov_system}). In this case the space dimension is
%two. The governing equations  are
%\begin{equation*}
%\left\{
%\begin{array}{c}
%\displaystyle h_{t}+\mathrm{div}\left( h\,\mathbf{U}\right) =0, \\
%\\
%\displaystyle\left( h\mathbf{U}\right) _{t}+\mathrm{{div}\,}\left(
%h\,\mathbf{U}\otimes \mathbf{U}+\frac{g\,h^{2}}{2}\ \mathbf{I}+h
%\,\sum\limits_{\alpha =1}^{2}\nabla \varphi _{\alpha }\otimes
%\nabla
%\varphi _{\alpha }\right) =\mathbf{0}, \\
%\\
%\qquad \displaystyle\frac{d\varphi _{\alpha }}{dt}=0,\ \ \alpha
%\in \{1,2\}.
%\end{array}
%\right.
%\end{equation*}
%The dissipation effects related to the wall friction can be added
%in the same way as in \cite{Whitham}:
%\begin{equation*}
%\left( h\mathbf{U}\right) _{t}+\mathrm{{div}\,}\left( h\,\mathbf{U}\otimes
%\mathbf{U}+\frac{g\,h^{2}}{2}\ \mathbf{I}+h\,\sum\limits_{\alpha
%=1}^{3}\nabla \varphi _{\alpha }\otimes \nabla \varphi _{\alpha }\right)
%=-C_{f}\mathbf{U}\left\Vert \mathbf{U}\right\Vert ,
%\end{equation*}%
%where $C_{f}$ is the Chezy coefficient associated with the flow friction.

\section{Conclusion}

The equations of fluid turbulent motions take three equations into account:
the equation of mass balance (\ref{System1}$_{1}$), the balance equation of
average momentum (\ref{System1}$_{2}$), and the Reynolds stress tensor
equation of evolution (\ref{System1}$_{3}$); this last equation has been the
object of our study. If the turbulent sources are neglected, the turbulent
fluid motion is a superposition of the mean motion and turbulent
fluctuations. The eigenvectors of the Reynolds stress tensor carry the
fluctuations associated with the mean flow deformation. The amplitudes of
turbulent deformations are defined by the eigenvalues of the Reynolds stress
tensor. The equations for the directions of turbulent fluctuations are
reminiscent of a gyroscopic type equation for the motion of a free rigid
body (Equation (\ref{driving system}$_{1}$)). The amplitude evolution of
turbulent deformations are determined by the diagonal values $\mu _{\alpha }$
of the mean rate of deformation tensor $\mathbf{D}$ expressed in the
eigenvector basis of the Reynolds stress tensor. The turbulence increases
with the time when $\mu _{\alpha }<0$, and decreases when $\mu _{\alpha }>0$%
. In the particular case of incompressible fluid motions we have $\mathtt{tr}%
\,\mathbf{D}=0$, and hence there always exists a direction in which the
turbulence is increasing while in other directions the turbulence is
decreasing. Over the past two decades great progress has been made in
understanding many aspects of the kinematics and dynamics of a wide variety
of turbulent flows as a result of access to the mean velocity gradient
tensor (see \cite{Marusic, Wallace}). Such an access could be used for an
experimental determining these eigenvalues. \newline
In the case of a small mean vorticity, a new approximate model admitting a
variational formulation is derived. %Such a model can
%be applied for studying turbulent shallow water flows.
\bigskip

\textbf{Acknowledgement} : We thank Professor Vladimir Teshukov who drew our
attention to the model of weakly sheared flows during his visit in Marseille
in the fall of year 2007, and Professor Boris Kolev for valuable discussions.


\begin{thebibliography}{10}
\bibitem[1]{Berdichevsky} Berdichevsky V., Variational Principles of
Continuum Mechanics, I. Fundamentals, Springer, Berlin, 2009.

\bibitem[2]{Berthon} Berthon C., Coquel F., H\'{e}rard J. M., Uhlmann M., An
approximate solution of the Riemann problem for a realisable second-moment
turbulent closure, Shock Waves, \textbf{11}, 245-269 (2002).

\bibitem[3]{Cahn} Cahn J. W., Hilliard J. E., Free energy of a nonuniform
system. I. Interfacial free energy, J. Chem. Phys., \textbf{28}, 258-67
(1958).

\bibitem[4]{Casal} Casal P., Capillarit\'{e} interne en m\'{e}canique des
milieux continus, C.\ R. Acad. Sci. Paris, \textbf{256}, 3820-3822 (1963).

\bibitem[5]{Casal_2} Casal P., La th\'{e}orie du second gradient et la
capillarit\'{e}, C. R. Acad. Sci. Paris, \textbf{274}, 1571--1574 (1972).

\bibitem[6]{Eglit} Eglit M. E., A generalization of the model of an ideal
compressible fluid, Journal of Applied Mathematics and Mechanics, \textbf{29}%
, 395--399 (1965).

\bibitem[7]{Debieve} Debi\`{e}ve J.-F., Gouin H., Gaviglio J., Momemtum and
temperature fluxes in a shock wave-turbulence interaction, in: Proceedings
of the ICHMT/IUTAM Symposium on the Structure of Turbulence and Heat and
Mass Transfer, Z. Zari$\mathrm{\acute{c}}$, Ed., pp. 277-296, Hemisphere
Publishing Corporation, London, 1982.

\bibitem[8]{Gouin} Debi\`{e}ve J.-F., Gouin H., Gaviglio J., Evolution of
the Reynolds stress tensor in a shock wave-turbulence interaction, Indian
Journal of Technology, \textbf{20}, 90-97 (1982).

\bibitem[9]{Dell'Isola} Dell'Isola F., Gouin H., Rotoli G., Nucleation of
spherical shell-like interfaces by second gradient theory: numerical
simulations, Eur. J. Mech. B/Fluids, \textbf{15}, 545-68 (1996).

\bibitem[10]{Marusic} Ganapathisubramani B., Longmire E. K., Marusic I.,
Experimental investigation of vortex properties in a turbulent boundary
layer, Phys. Fluids, \textbf{18}, 055105 (2006).

\bibitem[11]{Gavrilyuk2} Gavrilyuk S., Gouin H., A new form of governing
equations of fluids arising from Hamilton's principle, International Journal
of Engineering Science, \textbf{37}, 1485-1520, (1999).

\bibitem[12]{Gavrilyuk} Gavrilyuk S., Saurel R., Estimation of the
turbulence energy production across a shock wave, The Journal of Fluid
Mechanics, \textbf{549}, 131-139 (2006).

\bibitem[13]{Marsden} Marsden J. E., Ratiu T. S., Introduction to mechanics
and symmetry, Series in Applied Mathematics, \textbf{17}, Springer, Berlin,
1994.

\bibitem[14]{Pironneau} Mohammadi B., Pironneau O., Analysis of the
K-epsilon turbulence model, Research in Applied Mathematics, John Wiley \&
Sons, New York, 1994.

\bibitem[15]{Pope} Pope S. B., Turbulent flows, Cambridge University Press,
2005.

\bibitem[16]{Serrin} Serrin J., Mathematical principles of classical fluid
mechanics, in: Encyclopedia of Physics\emph{\/} \textbf{VIII/1}, pp.
125-263, Springer, Berlin, 1959.

\bibitem[17]{Truskinovsky} Truskinovsky L., Equilibrium phase boundaries,
Sov. Phys. Dokl., \textbf{27}, 551-553 (1982).

\bibitem[18]{Teshukov} Teshukov V. M., Gas dynamic analogy for vortex
free-boundary flows, Journal of Applied Mechanics and Technical Physics,
\textbf{48}, 303-309 (2007).

\bibitem[19]{Wallace} Wallace J. M., Twenty years of experimental and direct
numerical simulation access to the velocity gradient tensor : What have we
learned about turbulence?, Physics of Fluids, \textbf{21}, 021301 (2009).

\bibitem[20]{van_der_waals} van der Waals J. D., The thermodynamic theory of
capillarity under the hypothesis of a continuous density variation (1893).
Translated by J. S. Rowlinson, J. Stat. Physics, \textbf{20}, 197--244
(1979).

%\bibitem[21]{Whitham} Whitham G. B., Linear and Non-Linear Waves, Chap. 3, p.
%82, John Wiley \& Sons, New York, 1974.

\bibitem[21]{Wilcox} Wilcox D., Turbulence Modeling for CFD, DCW Industries,
1998.
\end{thebibliography}
\end{document}